\documentclass[conference]{vgtc}                          

\graphicspath{{figures/}{pictures/}{images/}{./}} 

\usepackage{times}                     

\usepackage{tabu}                      
\usepackage{booktabs}                  
\usepackage{lipsum}                    
\usepackage{mwe}                       
\usepackage{bibunits}
\defaultbibliography{template}

\usepackage{mathptmx}                  
\usepackage{comment}
\usepackage{enumitem}

\usepackage{subcaption}
\usepackage{xcolor}

\newif\ifshowchanges\showchangesfalse
\newcommand{\inserted}[1]{%
    \ifshowchanges
        \textcolor{red}{{}#1}%
    \else
        #1%
    \fi}
\newcommand{\deleted}[1]{%
    \ifshowchanges
    \else\fi}
\onlineid{0}

\vgtccategory{Research}

\vgtcinsertpkg

\title{GEMS - Guided Evolutionary Molecule Design for Sustainable Chemicals}

\author{Coelina Robinson\thanks{E-mail: \{crobinson, fweissbach, kjorner, melassad, humerc\}@ethz.ch} 
\and Franziska Weissbach
\and Kjell Jorner
\and Mennatallah El-Assady\
\and Christina Humer } %
\affiliation{\scriptsize ETH Zürich, Zurich, Switzerland}

\teaser{
    \centering
    \includegraphics[width=0.95\linewidth]{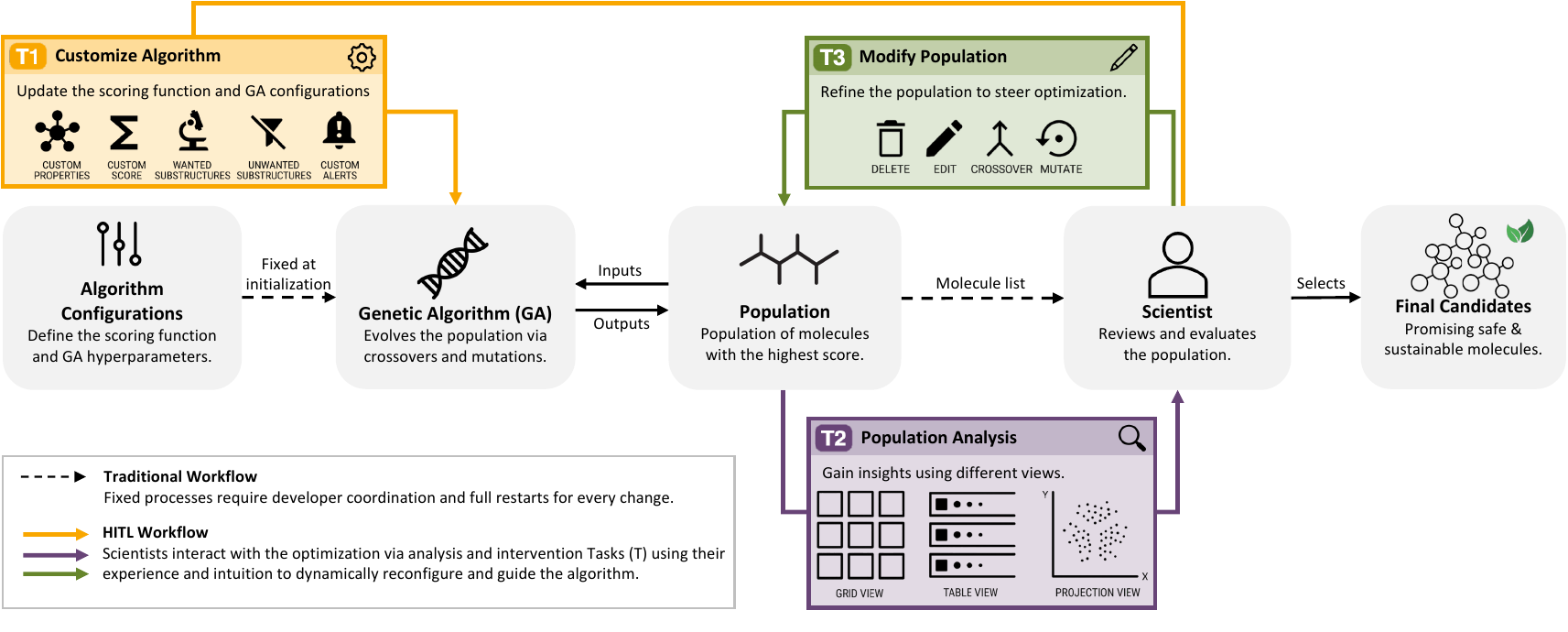}
    \vspace{3pt}
    \caption{A workflow comparison of traditional genetic algorithm pipelines \textit{vs.} our proposed Human-in-the-Loop (HITL) framework for sustainable molecule design. Unlike traditional linear, automated processes, our interactive workflow allows scientists to steer optimization, inspect candidates, and intervene in the evolutionary loop in real time.}
    \label{fig:Workflow}
}

\abstract{
    Designing safe and sustainable chemicals is critical to combat chemical pollution in our environment. 
    \deleted{Machine learning (ML)}\inserted{Computational and AI-assisted} methods have been developed to aid de novo molecule design. However, data on the environmental impacts of chemical compounds are sparse, resulting in low-fidelity \inserted{machine learning (}ML\inserted{)} oracles and unreliable candidate proposals.
    Furthermore, \inserted{many automated molecular design approaches} \deleted{generative ML models}rely on numerical scoring functions that cannot fully capture the nuanced chemical intuition of expert scientists required for real-world molecular design.
    \inserted{Instead, we present GEMS---an interactive visual analytics tool for human-in-the-loop molecular optimization that lets domain experts directly collaborate with an evolutionary genetic algorithm.}
    \inserted{Users continuously guide the search using domain knowledge through high-level, parametric modification of the scoring function alongside direct, granular control over molecule populations. GEMS requires no programming or expertise in ML or evolutionary optimization}.
    \deleted{A usage scenario demonstrates the system's application in designing sustainable antioxidant alternatives. In an interview session with domain scientists, we collected feedback on the usefulness of GEMS.}\inserted{A usage scenario demonstrates its application in designing sustainable antioxidant alternatives, and interviews with domain scientists provide feedback on its usefulness.}
} 


\usepackage[skins]{tcolorbox}

\definecolor{aubergine}{HTML}{684477}
\colorlet{auberginelight}{aubergine!40}
\definecolor{papaya}{HTML}{f9a700}
\colorlet{papayalight}{papaya!40}
\definecolor{olive}{HTML}{66852b}
\colorlet{olivelight}{olive!40}

\usepackage{overpic}

\usepackage{wrapfig2}

\makeatletter
\newtcbox{\kwColorBoxSpecial}[1][]{on line,fontupper=\footnotesize\sffamily\bfseries\small,boxrule=0.5pt,arc=2pt,coltext=white,colback=#1,colframe=#1,boxsep=0pt,left=1.5pt,right=1.5pt,top=1.5pt,bottom=1.5pt}
\newcommand{\kwSpecial}[2]{%
    \begin{kwColorBoxSpecial}[#2]%
    {#1}%
    \end{kwColorBoxSpecial}%
    \xspace%
}

\makeatletter
\newcommand{\defEntity}[2]{%
    \expandafter\gdef\csname entity@color@#2\endcsname{#1}%
    \ifx
        \protect\@typeset@protect
        \kwSpecial{\phantomsection\label{entity:#2}#2}{#1}%
    \else
        #2%
    \fi
}
\newcommand{\refEntity}[1]{%
    \@ifundefined{entity@color@#1}{%
        \PackageError{config.tex}{%
            Entity `#1' referenced before definition%
        }{%
            Use \string\defEntity\{<color>\}\{#1\} before calling
            \string\refEntity\{#1\}.%
        }%
    }{%
        \ifx
            \protect\@typeset@protect
            \hyperref[entity:#1]{\kwSpecial{#1}{\csname entity@color@#1\endcsname}}%
        \else
            #1%
        \fi
    }%
}
\makeatother

\newcommand{\dimension}[5]{
    \begin{tcolorbox}[
        fonttitle=\bfseries,
        coltitle=black,
        colbacktitle=#2!40,
        colback=#2!20,
        colframe=#2,
        title=#1, 
        after skip=0.35em,
        left=6pt, 
        right=3pt, 
        top=3pt, 
        bottom=3pt,
        boxsep=0pt,
        colbacklower=#2!5,
        middle=0.4em,
        toptitle=6pt, 
        bottomtitle=4pt,
        arc=0mm,           
        boxrule=0.0mm,     
        leftrule=0.7mm     
    ]
    #3
    \end{tcolorbox}%
}

\begin{document}



\maketitle
 
\begin{bibunit}
\section{Introduction}

Chemical pollution poses a critical threat to human health and our Earth's ecosystems\deleted{~\cite{richardson_earth_2023}}. Rapid chemical production has outpaced our ability to assess compound safety, necessitating a shift from reactive testing to proactive, safe-and-sustainable-by-design (SSbD) methodologies~\cite{fenner_need_2021}. 
Designing alternative molecules that retain the desirable functional properties of existing pollutants, such as PFAS, antioxidants, or refrigerants~\cite{brunn_pfas_2023,goldszal_discovery_2025,seller-brison_hazard_2026}, while being safer and more sustainable, is therefore an important yet challenging task. Environmental safety and sustainability are difficult to predict due to limited data availability\inserted{~\cite{wang_addressing_2025}}, and their assessment still relies on scientists' expertise and intuition\inserted{. Such expertise spans synthetic feasibility and greenness, persistence estimation~\cite{posthumus_external_2005}, and recognition of instability and reactivity, such as transformation products and interactions with organisms~\cite{jayasekara_assessing_2021}}.

Existing \deleted{ML}\inserted{computational and AI-assisted molecular design} methods aim to support this process, but are often not easily accessible to domain scientists~\cite{vieira_imesc_2025} because they require programming knowledge or developer support to use and adapt them to scientists' needs. Furthermore, current models remain limited by the accuracy of available oracles and the lack of any oracles for many relevant and often context-dependent properties~\cite{jain_gflownets_2023}. As a result, effective molecular design still requires the supervision of domain experts to guide and refine the process~\cite{he_collaborative_2025, nahal_human---loop_2024}.

To address this gap, we developed an interactive visual tool that enables human experts to integrate their \deleted{expertise}\inserted{intuition for the relationship between molecular structure and diverse properties} with a generative system for molecular design without requiring technical knowledge \inserted{of ML models or evolutionary optimization methods}. The platform allows users to explore and analyze generated molecules, steer the genetic algorithm, and iteratively refine results throughout the evolutionary process. Our approach combines efficient computational optimization with the generalizable, intuitive knowledge of human experts in an interactive workflow.

\section{Background and Related Work}
We provide an overview of genetic algorithms and methods for interactive molecular visualization and exploration, human-guided optimization, and their application to environmental chemistry.

\subsection{Genetic Algorithms}
\deleted{Genetic algorithms (GAs) are} \inserted{A genetic algorithm (GA) is an} evolutionary optimization method\deleted{s} inspired by natural selection\inserted{. GAs}\deleted{that} iteratively evolve a population of candidates~\cite{mitchell_introduction_1996}. In each iteration, the fittest candidates (\textit{i.e.}, highest scoring according to a scoring function) are selected and modified by crossover and mutation operations. Crossover recombines molecular substructures of two parent candidates to produce novel offspring that inherit features from both parents. Mutation introduces local structural changes to individual candidates, \deleted{ensuring continued exploration of}\inserted{exploring} the search space. These operations can be handled stochastically or by integrating other models, such as LLMs~\cite{wang_efficient_2025}.

GAs have proven to be strong baselines for molecular generation, often outperforming more complex generative models on standard benchmarks~\cite{tripp_genetic_2023}. GEMS builds on graph-based GA operators that perform crossover and mutation directly on molecular graphs~\cite{brown_guacamol_2019,jensen_graph-based_2019}. 

\subsection{Interactive Molecular Visualization and Exploration}
Several visual analytics systems have been developed to support chemists in exploring and interpreting molecular data. A common approach is to use dimensionality reduction techniques to efficiently process the high-dimensional space of molecular features, enabling visual assessment of structural diversity and clustering\deleted{~\cite{sabando_chemva_2021}}~\cite{cihan_sorkun_chemplot_2022,humer_cheminformatics_2022,humer_cime4r_2024}. To support systematic comparison and filtering across molecular properties, multi-attribute table views are a popular choice~\cite{gratzl_lineup_2013,kale_chemograph_2023}. Beyond exploration, recent systems have incorporated model explanations~\cite{humer_cheminformatics_2022,humer_cime4r_2024,rodriguez-perez_interpretation_2020} and uncertainty visualization~\cite{humer_cime4r_2024} to make data-driven predictions more transparent, and Menke et al.~\cite{menke_metis_2024} have introduced a GUI to collect expert feedback for generative chemistry models. 

Our work takes inspiration from existing research in the field and specifically CIME~\cite{humer_cheminformatics_2022} in using a combination of structure, table, and projection views, but differs in a key aspect: rather than analyzing precomputed results or existing datasets, our tool enables users to directly steer a generative optimization process through iterative interaction with a genetic algorithm.

\subsection{Human-Guided Molecular Optimization}

Machine learning has become a central tool for molecular design, with approaches ranging from variational autoencoders and reinforcement learning to diffusion models to generate novel candidate structures~\cite{anstine_generative_2023}. However, current models still struggle to capture the nuanced, experience-driven judgments that chemists bring to molecular evaluation\deleted{of, for example,}, such as synthetic feasibility or functional suitability, which are difficult to encode in a computational scoring function~\cite{choung_extracting_2023,jain_gflownets_2023} and necessitate trial-and-error adjustments~\cite{fujii_chemtsv3_2025}. This \deleted{and similar problems in other fields have}\inserted{has} motivated efforts toward human-in-the-loop generative modeling~\cite{mosqueira-rey_human---loop_2023}, \deleted{in which }\inserted{where }human experts guide model behavior \deleted{. The guidance can target different points in the modeling pipeline and employ various techniques, such as}\inserted{via techniques like} preference learning~\cite{choung_extracting_2023,menke_metis_2024,sundin_human---loop_2022} or active learning~\cite{nahal_human---loop_2024}. \deleted{Furthermore, different model steering techniques have been developed across domains, from interactive visualization of optimization campaigns in chemistry~\cite{humer_cime4r_2024} and explanation-driven data reconfiguration in healthcare~\cite{bhattacharya_explanatory_2024}, to the use of human interactions for inference-time biasing of generative sampling policies in robotics~\cite{wang_inference-time_2025}.}\inserted{Existing steering approaches, such as interactive visualizations of optimization campaigns in chemistry~\cite{humer_cime4r_2024}, typically mediate human input through a model---expert feedback updates a scoring or surrogate function, which then only indirectly drives the optimization.}
\deleted{A common thread in these approaches is that human input is mediated through a model---experts provide feedback that is integrated into a scoring or surrogate function, which then indirectly drives the optimization.}

Our tool takes a different approach: instead of channeling expert knowledge through a scoring function, it provides chemists with direct, continuous control over a \deleted{genetic algorithm}\inserted{GA} \inserted{acting as an evolutionary optimizer}, transforming the expert from \deleted{an oracle queried by the system}\inserted{a queried oracle} into an active collaborator who shapes the search at every iteration.

\subsection{Safe and Sustainable Chemical Design}
While human-in-the-loop approaches have begun to appear in drug discovery~\cite{menke_metis_2024,nahal_human---loop_2024,sundin_human---loop_2022}, our tool focuses on sustainability-oriented design. The benefit of interactivity is particularly important in SSbD because computational oracles for toxicity and persistence are less mature than for, \textit{e.g.}, binding affinity, \inserted{which is }central to medicinal chemistry\inserted{, and it is important to quickly identify and remedy reward hacking}. The first generative modeling approaches to designing environmentally friendly chemicals have made progress~\cite{wang_paretogen_2026,yang_molecular_2024}, but they remain constrained by property predictors trained on sparse data that often fail to generalize to novel structures and have no options to influence the generation process.

\section{Goals and Core Tasks}
The main goal of this tool is to support scientists in collaborating with a GA to design new, safe, and sustainable chemicals. It enables users to fine-tune the system to their specific use case without support from developers and to integrate domain knowledge and expert intuition to mitigate ML drawbacks. Based on discussions with domain scientists, we identified the following core tasks:

\dimension{\defEntity{papaya}{T1} Customize Algorithm}{papaya}{
Scientists need to be able to run a genetic algorithm for molecular design out of the box, while retaining the ability to customize it for different research goals.
}{}{}
\begin{itemize}[noitemsep]
    \item[\defEntity{papaya}{T1.1}] \textbf{Configuring key parameters} such as population size and number of generation cycles without the need for programming or computational modeling knowledge.
    \item[\defEntity{papaya}{T1.2}] \textbf{Customizing the scoring function} and updating it throughout the evolutionary process to reflect specific research goals and changing knowledge on how to reach these goals.
    \item[\defEntity{papaya}{T1.3}] \textbf{Defining structural constraints} to incorporate expert knowledge on desirable or undesirable substructures in the search.
\end{itemize}

\dimension{\defEntity{aubergine}{T2} Population Analysis}{aubergine}{
Scientists need to analyze the evolving population to identify promising design directions and diagnose potential issues in the optimization process. Based on insights from this analysis, scientists can make informed decisions on how to adjust the scoring function or the population.
}{}{}
\begin{itemize}[noitemsep]
    \item[\defEntity{aubergine}{T2.1}] \textbf{Investigating candidates} in detail to help experts learn and adapt their knowledge of the search domain. Users should be able to investigate candidate features, structures, and scores and compare them to other candidates.
    \item[\defEntity{aubergine}{T2.2}] \textbf{Exploring the population} to assess whether the genetic algorithm is appropriately balancing exploration and exploitation of the search space. 
\end{itemize}

\dimension{\defEntity{olive}{T3} Modify Population}{olive}{
Scientists need to be able to manually adapt the population and modify candidates. This is in addition to the higher-level configuration of the scoring function and allows for a fine-grained way of steering the optimization process by utilizing users' internalized expertise.
}{}{}
\begin{itemize}[noitemsep]
    \item[\defEntity{olive}{T3.1}] \textbf{Removing molecules} from the current population allows users to exclude undesirable or implausible candidates as input for the next GA generation. 
    \item[\defEntity{olive}{T3.2}] \textbf{Guiding mutation and crossover} operations to allow for a custom treatment of particularly promising candidates.
    \item[\defEntity{olive}{T3.3}] \textbf{Editing structures} of molecules to enable fine-grained control of the candidates beyond the generations of the GA.
\end{itemize}

\begin{figure}[htbp]
    \centering
    \begin{subfigure}[c]{0.615\linewidth}
        \centering
        \includegraphics[width=\linewidth]{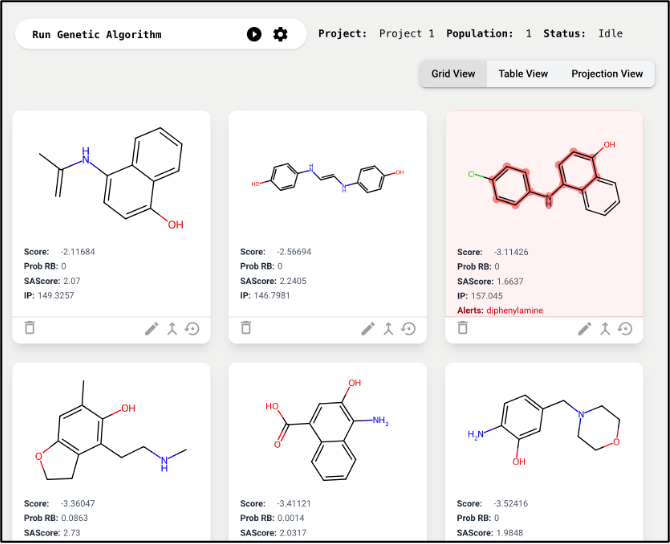}
        \caption{Grid View.}
        \label{fig:grid}
    \end{subfigure}
    \hfill
    \begin{minipage}[c]{0.35\linewidth}
        \centering
        \begin{subfigure}{\linewidth}
            \centering
            \includegraphics[width=\linewidth]{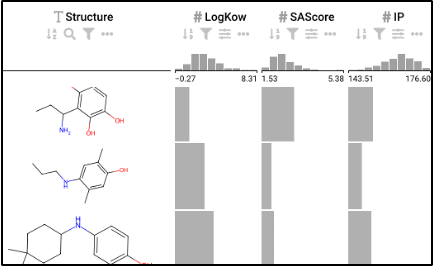}
            \caption{Table View.}
            \label{fig:table}
        \end{subfigure}
        
        \begin{subfigure}{\linewidth}
            \centering
            \includegraphics[width=\linewidth]{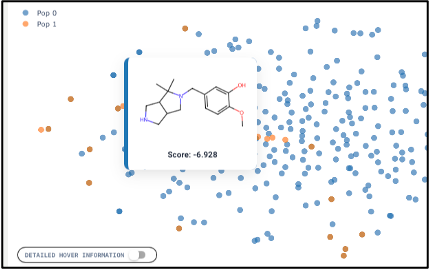}
            \caption{Projection View.}
            \label{fig:projection}
        \end{subfigure}
    \end{minipage}

    \caption{The tool provides three views for exploration: (\subref{fig:grid}) the Grid View, (\subref{fig:table}) the Table View, and (\subref{fig:projection}) the Projection View.}
    \label{fig:views}
\end{figure}

\section{System Design}
\deleted{In this section, we first provide an overview of the platform and walk through the underlying workflow. We then discuss the individual components in more detail, explaining how they assist users in performing their specific tasks\footnote{Code available at: \url{https://github.com/ETH-IVIA-Lab/gems}.}.}
\inserted{This section outlines the platform's workflow and details how individual components assist users in performing their specific tasks\footnote{Code available at: \url{https://github.com/ETH-IVIA-Lab/gems}.}}.

\subsection{Overview \& Workflow}

Figure~\ref{fig:Workflow} illustrates a traditional genetic algorithm workflow compared to the human-in-the-loop workflow enabled by GEMS. In the traditional setting, a developer programs the scoring function and configures the algorithm, after which the GA runs iteratively and returns the resulting population to the scientist, leaving little room for expert input or adjustment during the process. Human feedback is limited to inspecting the final output and adjusting the algorithm's initial configuration accordingly.

With GEMS, scientists can define and modify the scoring function themselves through an intuitive interface without requiring developer support. Crucially, both the scoring function and the algorithm's hyperparameters can be updated at any point during the process without restarting, allowing the search to be continuously steered between generations as new insights emerge (\refEntity{T1}). The resulting molecules are visualized through three complementary views, a grid view, a table view, and a projection view (Figure~\ref{fig:views}), each supporting different analytical tasks (\refEntity{T2}). Through these interfaces, users can inspect and analyze the population, intervene directly by deleting undesirable candidates, mutating or crossing over promising molecules, or manually editing structures (\refEntity{T3}). Users can then re-run the GA on the modified population, closing the human-in-the-loop cycle and iteratively guiding the search toward desired molecular properties. This functionality is delivered via a lightweight web application to eliminate installation barriers, ensuring domain scientists can access the system regardless of their local computational environment.

\subsection{Algorithm Configurations}

\begin{figure}[htbp]
    \centering
    \begin{subfigure}[b]{0.35\linewidth} 
        \centering
        \includegraphics[height=4.4cm, keepaspectratio]{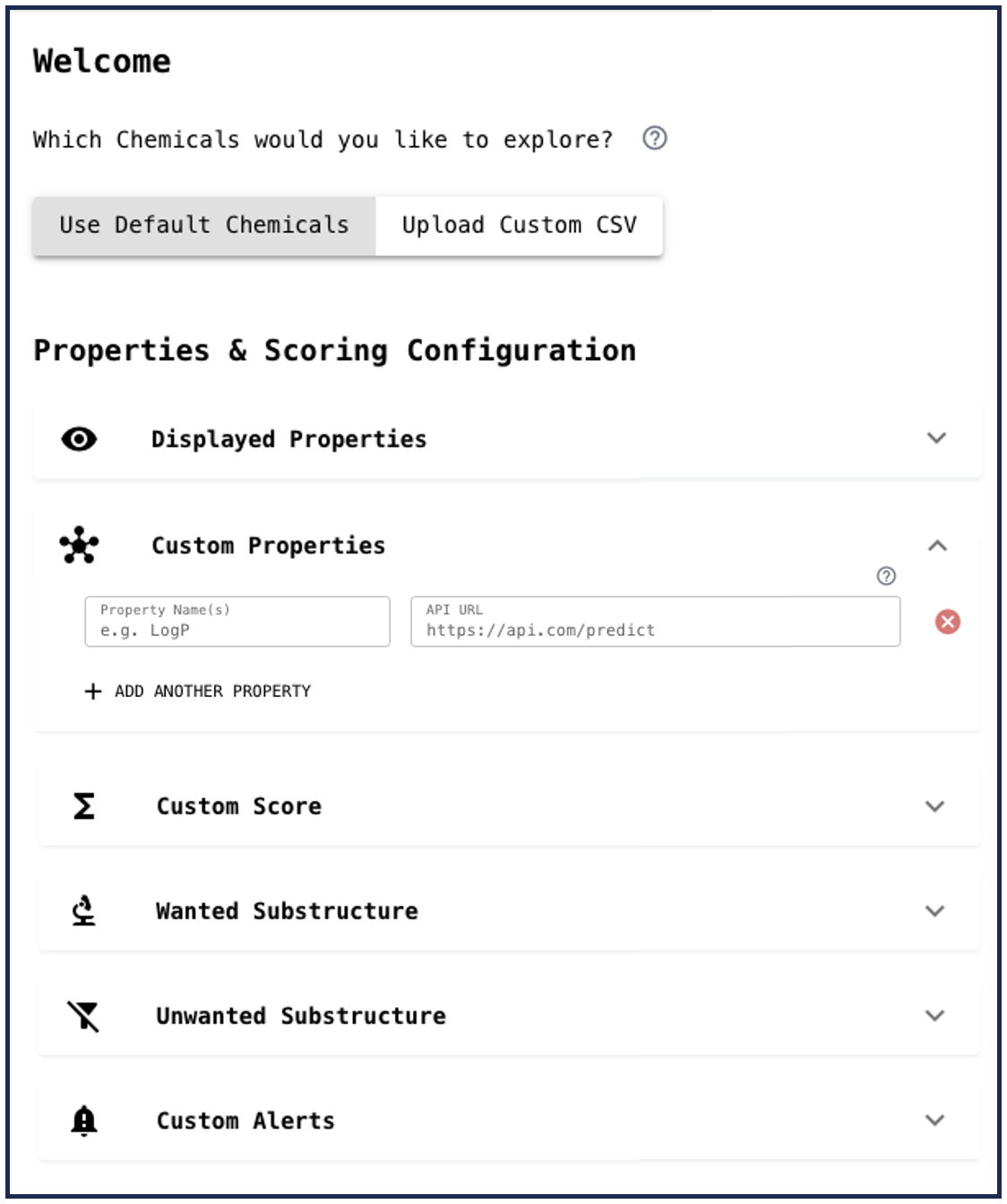}
        \caption{} 
        \label{fig:config1}
    \end{subfigure}
    \hfill 
    \begin{subfigure}[b]{0.55\linewidth}
        \centering
        \includegraphics[height=4.4cm, keepaspectratio]{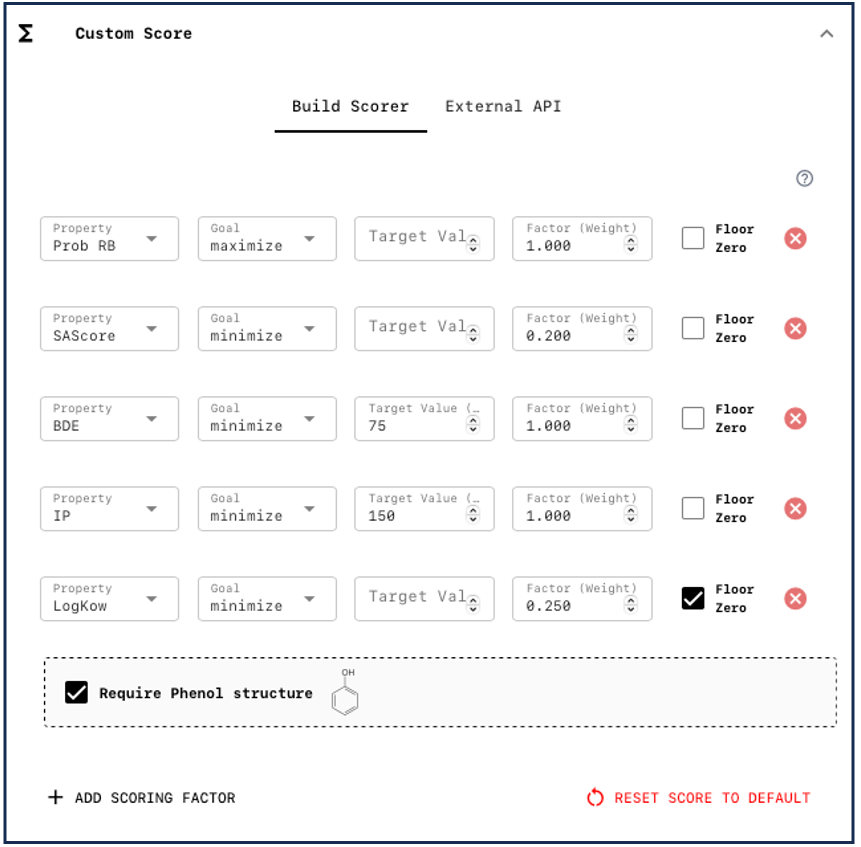}
        \caption{}
        \label{fig:config2}
    \end{subfigure}
    
    \caption{The configuration interface (\subref{fig:config1}) allows users to upload a dataset, customize attributes and the scoring function. The expanded custom property tab shows API integration. The score editor (\subref{fig:config2}) allows viewing and modifying the default scoring function.}
    \label{fig:config}
\end{figure}

To satisfy \refEntity{T1.1}, the GA supports a single-click launch with sensible defaults, while still allowing advanced users to manually tune hyperparameters like population size. When launched, users are presented with an overview of configurable components, and the option to either upload their own dataset of molecules or use a predefined sample from the PubChem compound database (Figure~\ref{fig:config1}). Users are provided with a graphical scoring editor (Figure~\ref{fig:config2}) that lets scientists define and update scoring functions without programming knowledge or developer support. Users are presented with a default scoring function that balances functionality, synthesizability, and environmental impact, which can serve as a starting point or be fully reconfigured by adjusting property weights or adding or removing terms entirely (\refEntity{T1.2}).

To give users full flexibility in scoring and allow them to include trusted external tools, external APIs can be integrated into the scoring function, extending it with custom molecular properties beyond the defaults (Figure~\ref{fig:config1}). Substructure-based rewards and penalties address \refEntity{T1.3} by allowing experts to encode structural intuition directly, for example, penalizing known persistent pollutant scaffolds without translating them into explicit scoring terms. Alerts can also visually flag substructures in the population view (Figure \ref{fig:grid}), keeping expert knowledge visible throughout the process rather than only at configuration time.

Critically, none of these configurations are locked after initialization. Users can analyze the population (\refEntity{T2}) throughout the process to update the scoring function, weights, and hyperparameters. This is a deliberate change from traditional GA pipelines, where the scoring function is fixed before execution. Here, real-time reconfiguration allows the search trajectory to respond to users' observations as the population evolves.

\subsection{Exploration and Analysis Views}

The tool provides three distinct interfaces (Figure~\ref{fig:views})\deleted{ designed to support varied analytical needs. The Grid View addresses the need for rapid visual screening, users can spot obvious undesirable candidates at a glance.}. \inserted{The Grid View enables rapid visual screening. By embedding alerts directly into the 2D molecular drawings, chemists can instantly spot and remove undesirable candidates before running the next GA cycle.}
\deleted{The Table View serves a different analytical need, when users want to analyze, sort and filter by multiple attributes and identify candidates that score critically low on any single criterion (\refEntity{T2.1}).} 
\inserted{The Table View allows users to analyze, sort, and filter candidates across multiple properties. This design is valuable for SSbD evaluation, where chemists must balance conflicting objectives like maximizing chemical activity while minimizing environmental persistence, quickly filtering out candidates that achieve a high score but fail a single, critical safety threshold (\refEntity{T2.1}).}
Finally, the Projection View addresses \deleted{a third distinct need: assessing the }population-level structure. \deleted{By }Embedding molecules into 2D space using Morgan fingerprints~\cite{rogers_extended-connectivity_2010} \inserted{allows users to identify search convergence} \deleted{users can identify whether the search is converging }or locate structurally similar neighbors to a promising candidate to better understand the effect of specific structural changes (\refEntity{T2.2}).
Together, these views enable a multi-faceted understanding of complex search spaces (\refEntity{T2}) \inserted{and support decision-making}. 

\subsection{Interactive Steering and Refinement}

\begin{figure}[htbp]
    \centering
    \begin{subfigure}{\linewidth}
        \centering
        \begin{minipage}[c]{0.05\linewidth} 
            \caption{}
            \label{fig:modifier-crossover}
        \end{minipage}
        \begin{minipage}[c]{0.86\linewidth} 
            \centering
            \includegraphics[width=\linewidth]{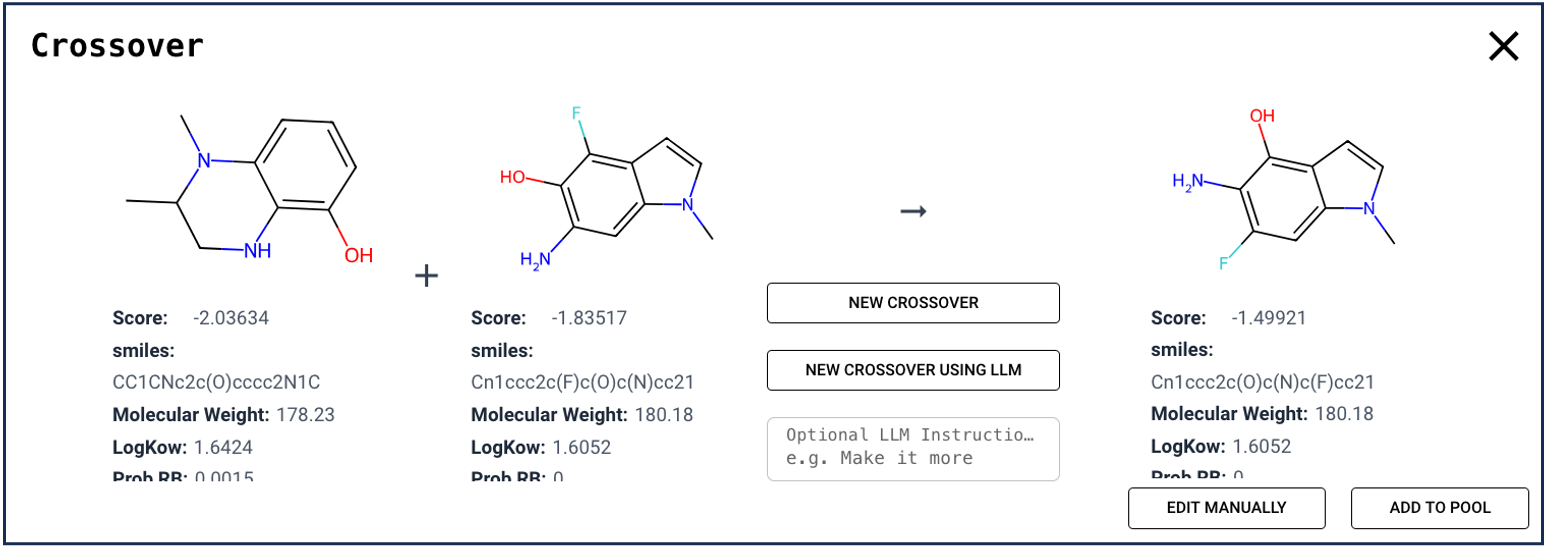}
        \end{minipage}
    \end{subfigure}

    \begin{subfigure}{\linewidth}
        \centering
        \begin{minipage}[c]{0.05\linewidth}
            \caption{}
            \label{fig:modifier-edit}
        \end{minipage}
        \begin{minipage}[c]{0.86\linewidth}
            \centering
            \includegraphics[width=\linewidth]{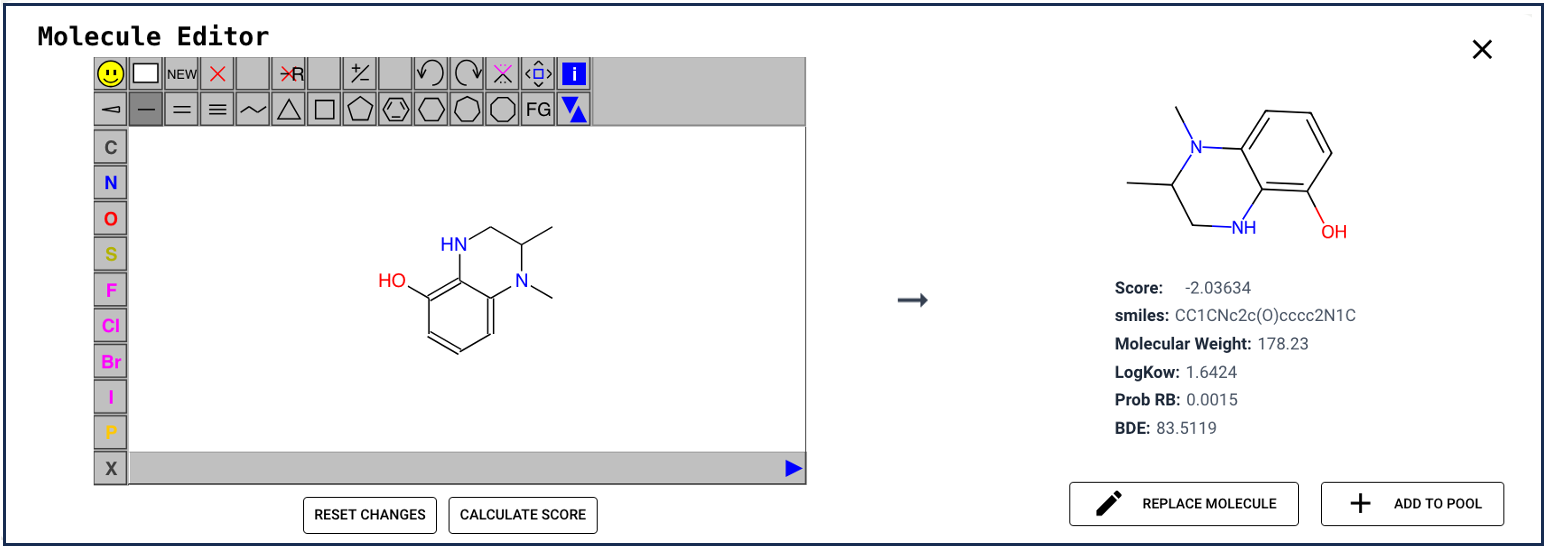}
        \end{minipage}
    \end{subfigure}

    \caption{Interaction interfaces for manual population steering: (\subref{fig:modifier-crossover}) crossover and (\subref{fig:modifier-edit}) direct structure editing.}
    \label{fig:modifier}
\end{figure}

Effective molecular design requires intervention at different levels of granularity depending on what the user observes.  Deletion enables fine-grained filtering of undesirable candidates (\refEntity{T3.1}), while manual mutation and crossover (Figure~\ref{fig:modifier-crossover}) allow users to selectively apply evolutionary operators to promising candidates rather than leaving the recombination entirely to the GA (\refEntity{T3.2}). For precise structural changes, an integrated drawing tool~\cite{bienfait_jsme_2013} (Figure~\ref{fig:modifier-edit}) enables edits beyond autonomous GA capabilities (\refEntity{T3.3}). Users can also prompt an LLM in natural language to mutate or cross over molecules, which is useful when users have an abstract intuition that can be translated into various concrete structures. After any combination of interventions, the GA can be run on the modified population, closing the human-in-the-loop cycle.

\section{Usage Scenario}
To illustrate how our tool\footnote{Demo available at: \url{https://gems.ivia.ch/}.} \deleted{supports scientists }\inserted{can be used by scientists}, we walk through the following scenario: 
Synthetic antioxidants are widely used in consumer products such as food packaging, skincare, and plastics. However, many have been flagged for environmental persistence and ecotoxicity, creating an urgent need for safer alternatives~\cite{seller-brison_hazard_2026}. We illustrate this challenge through a scenario in which a chemist, Alice, uses our tool to identify novel antioxidant candidates that are both effective and environmentally friendly.
 
Alice begins by uploading a curated dataset of diverse phenolic compounds, leveraging the fact that phenol groups are the key component that makes many natural and synthetic antioxidants effective~\cite{seller-brison_hazard_2026}. Her curated starting population ensures structural diversity while anchoring the search in a chemically meaningful space. She then configures the scoring function to balance four objectives equally: (1) antioxidant activity, (2) toxicity, (3) environmental persistence, and (4) synthesizability (Figure~\ref{fig:config2}). This reflects the multi-objective nature of safe-by-design molecular discovery. Drawing on domain knowledge, she additionally specifies structure-based penalties for tert-butyl substituents, which are common in commercial antioxidants but associated with environmental persistence~\cite{seller-brison_hazard_2026}, as well as for bisphenol A and estrogen-like scaffolds that are linked to endocrine disruption~\cite{rubin_bisphenol_2011}, meaning hormone interference. Furthermore, she creates an alert for diphenylamines, a motif that can degrade into toxic primary aromatic amines \cite{seller-brison_hazard_2026}, allowing case-by-case judgment rather than penalty.

After saving her settings, Alice inspects the initial population in the Grid View, sorted by the scoring function, which shows each molecule's properties and score (Figure~\ref{fig:grid}). She removes molecules flagged with red alerts, and then switches to the Table View (Figure ~\ref{fig:table}) to sort by individual properties and delete candidates that perform poorly on any single criterion, ensuring a clean starting population. She runs the GA and observes which structural motifs persist across generations. She notices that halogenated phenol derivatives consistently exhibit high antioxidant activity, \deleted{and their low biodegradability does not compensate for this.}\inserted{and the penalty for their low biodegradability is insufficient to outweigh this advantage.} In response, she increases the weight of biodegradability in the scoring function and uses the LLM-guided mutation feature to suggest alternative substituents for a selected candidate. Afterwards, she runs more GA cycles.

\deleted{In the final review}\inserted{Lastly}, she uses the Projection View (Figure~\ref{fig:projection}) to assess the diversity of the evolved population. While later generations occupy a higher-scoring but increasingly narrow region of chemical space, she identifies compounds at the periphery that combine promising activity and safety scores with low synthesizability. She manually edits these structures to \deleted{improve their tractability for synthesis}\inserted{simplify their synthesis} and removes near-duplicate compounds in regions of high structural similarity. The result is a shortlist of diverse, high-scoring antioxidant candidates that reflects both algorithmic optimization and expert judgment for later synthesis and testing, \deleted{an outcome that is difficult for either the GA or the chemist alone to achieve}\inserted{demonstrating the potential of integrating human insight directly into the evolutionary loop}.

Another scientist, Bob, focuses on refrigerants for heat pumps; he uses the tool by adapting the default scoring function to his objectives. Not all required properties are available for this use case by default; however, Bob can extend the system via custom API endpoints. Bob adds an endpoint to compute the global warming potential and the cooling capacity~\cite {goldszal_discovery_2025}. He then defines whether each property should be minimized, maximized, or targeted, and assigns corresponding weights. This scenario showcases how the tool can be flexibly adapted to a wide range of research needs.

\section{\inserted{Evaluation} \deleted{User Feedback}}

\deleted{The purpose of the interview session was to gain a general understanding of the tool's usability and design. We collected feedback from two domain scientists with little to no experience with generative models in chemistry. Each participant completed two tasks: optimizing the default scoring function and steering toward simpler, more biodegradable antioxidant structures. Then they completed a usability and user experience survey; responses are provided in the supplementary material.}
\inserted{To assess the tool’s usability and design, we collected feedback from two domain scientists with minimal experience in generative chemistry models. Each participant completed two tasks: optimizing the default scoring function and steering toward simpler, more biodegradable antioxidant structures, followed by a usability survey (responses are in the supplementary material).
}

The feedback was overall positive. The tool was considered useful (4.5/5) and well-designed (4/5), but received a passable score for intuitiveness (3.5/5). The configuration interface and constraint features (\textit{e.g.}, exclude substructures) were highly valued. Participants favored the Grid View for inspection and the Table View for filtering and sorting. \inserted{The Projection View was used less due to cross-view tracking difficulties. This could be solved by a better linking strategy of the views such that users can see their selections and modifications across views.} Mutation, crossover, and manual editing were considered valuable for making targeted structural changes. Participants' main difficulty was understanding the genetic algorithm, which required explanation; they suggested adding more onboarding and guidance. 
Nevertheless, they reported enjoying the interaction and would use GEMS again for their research (4.5/5).

\inserted{As preliminary evidence for the benefits of expert steering, we compared the top-100 candidates produced by a standard GA run with those of an expert-guided generation. While the standard GA achieved higher scores, the expert-guided candidates were more diverse and markedly more often solvable by computational retrosynthesis (65 vs. 32 of 100). A group of three environmental scientists expressed a slight preference for the expert-evolved candidates in a blind assessment but noted high uncertainties in their assessment due to a lack of experimental data for similar compounds. Details can be found in the Appendix.}

\section{Conclusion \& Future Work}

We present GEMS, a human-in-the-loop visual tool for sustainable molecular design that combines algorithmic exploration with direct expert intervention. Unlike in traditional GA workflows, the domain scientist is an active collaborator with direct control over the evolutionary process, enabling them to inspect, edit, and steer evolving populations. Early feedback indicated that participants would use GEMS in their research.

Future work will include adding an onboarding option, conducting an extended user study examining steering strategies, and leveraging implicit feedback from user interactions to train a preference model that reflects expert intuition. Furthermore, displaying scoring explanations could help users calibrate trust in the underlying surrogate models.

\acknowledgments{
We thank Stefan Schmid, Cleo Soldini, and Bernadette Mederer for participating in the user study. This work was funded by the Swiss National Science Foundation (SNSF) within the Sinergia project CRSII\_222712, as well as the ETH AI Center postdoctoral fellowship awarded to Christina Humer.
}
\\

\defaultbibliographystyle{abbrv-doi-hyperref-narrow}

\putbib
\end{bibunit}
    
\clearpage
\appendix
\crefalias{section}{appendix} 

\section{Appendix}
\begin{bibunit}
\subsection{Detailed description of the evaluation of the value of expert steering}
\label{appendix:evaluation}
While a controlled efficacy study is beyond the scope of this short paper, we provide preliminary, illustrative evidence for the benefit of expert steering by comparing the 100 highest-ranked compounds of a single standard GA run and a single expert-steered run using the GEMS platform. Both used the same starting population sampled from PubChem and the same initial scoring function aimed at producing safe, sustainable and functional antioxidants. The standard GA run was terminated after the highest achieved score of the population did not increase for 20 generations following the guidance of Jensen~\cite{jensen_graph-based_2019}, while the termination of the interactive generation process was left up to the expert. The baseline GA run produced higher scoring molecules when comparing both sets of candidates with the same default scoring function. However, this is to be expected due to dynamic adjustments made to the scoring function in the interactive generation setting. Furthermore, the initial scoring function is not the sole measurement that needs to be taken into account. In the following, we analyze the two candidate lists based on diversity, synthesizability, expert preference, and structural patterns---important factors for the successful design of an SSbD chemical. The two candidate lists can be found in the supplementary material.

\subsubsection{Diversity}
To assess diversity of the shortlists, the number of unique Murcko scaffolds \cite{bemis_1996_PropertiesKnownDrugs} and the mean pairwise Tanimoto distance within each candidate set were calculated using rdkit \cite{RDKitOpensourceCheminformatics}. The Tanimoto distance was calculated based on Morgan fingerprints with a radius of 2 and 2048 bits \cite{rogers_extended-connectivity_2010}. The expert-guided shortlist is more diverse considering any of these metrics, but the difference is most prominent for the number of unique Murcko scaffolds.

\begin{table}[h!]
    \centering
    \caption{The number of unique Murcko scaffolds and the mean pairwise Tanimoto distance are used to assess diversity of the candidate lists, where higher numbers correspond to higher diversity.}
    \begin{tabular}{lcc}
    \toprule
         & \# Murcko scaffolds & mean Tanimoto distance \\
         \midrule
         Baseline & 3 & 0.724 \\
         Expert-steered & 20 & 0.766 \\
         \bottomrule
    \end{tabular}
    \label{tab:placeholder}
\end{table}

\subsubsection{Synthesizability}
While synthesizability is considered in the default scoring function in the form of the heuristic SAScore~\cite{ertl_2009_EstimationSyntheticAccessibility}, a low SAScore is not a guarantee for a molecule to be synthesizable. Retrosynthesis tools like AiZynthFinder \cite{genheden_aizynthfinder_2020} can serve as a higher-accuracy oracle but are usually prohibitively computationally expensive to integrate into generative models for molecular design. We ran AiZynthFinder with its default configurations to find synthesis routes for both shortlists and found that 65 of the 100 interactively evolved candidates can be solved for a valid synthesis route, while only 32 of the baseline candidates can.

\subsubsection{Expert preference}
The two shortlists were furthermore shown to three environmental scientists to judge their suitability for the development of safe and functional antioxidants. The experts noted potential problems in terms of stability for both sets of compounds and a high uncertainty for any judgment due to the lack of similar compounds with available experimental data quantifying hazards and function but expressed a slight preference for the human-guided shortlist.

In light of the emerging chemical reasoning capabilities of frontier LLMs \cite{mirza_2025_FrameworkEvaluatingChemical}, we prompted Claude Opus 4.8 to evaluate the two shortlists blindly for safety and functionality in three replicates. In all three answers, the model preferred the human-guided shortlist referencing typical toxicity alerts, concerns about synthesizability and instability for the baseline population. The full interactions can be found in the Supplementary Material.

\subsubsection{Structural patterns}
The highest ranked compounds from both shortlists are shown in Figure~\ref{fig:compounds} below.

\begin{figure}[htbp]
    \centering
    
    \begin{subfigure}[b]{0.49\linewidth} 
        \centering
        \includegraphics[height=2cm]{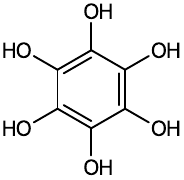}
        \caption{highest scoring compound in the baseline shortlist} 
        \label{fig:baselinecompound}
    \end{subfigure}
    \hfill 
    \begin{subfigure}[b]{0.49\linewidth}
        \centering
        \includegraphics[height=2cm]{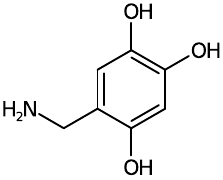}
        \caption{highest scoring compound in the expert-steered shortlist}
        \label{fig:expertcompound}
    \end{subfigure}
   
    \caption{}
    \label{fig:compounds}
\end{figure}

They illustrate some of the structural trends separating the two shortlists. The baseline set consists overwhelmingly of densely hydroxylated single-ring phenols of which the compound in Figure~\ref{fig:baselinecompound} is the most extreme example. Extensive hydroxylation can lower the phenolic O–H bond-dissociation enthalpy, which is used as a proxy for antioxidant activity, but such electron-rich polyols may also be susceptible to autoxidation and pro-oxidant behavior and tend to be synthetically demanding. They are frequently accompanied by known toxicity alerts, including catecholamine-type amines (49\%), hydrazines (9\%), and nitrogen–oxygen motifs (9\%). These features are consistent with reward hacking, whereby the optimization exploits the scoring function by maximizing hydroxylation and adding electron-rich groups that inflate the activity term while degrading properties the oracle captures poorly. The expert-steered set is more moderate, with lower hydroxylation and an absence of the aforementioned reactive nitrogen motifs. Its highest-ranked candidate (see Figure~\ref{fig:expertcompound}), a trihydroxybenzene bearing a single catechol and an aminomethyl group, may retain radical-scavenging capacity while avoiding the extreme hydroxylation and structural alerts of the baseline, though a subset of the expert-steered candidates instead consist of biaryl polyphenols (15\%), which may show larger numbers of hydroxyl groups.

\putbib

\end{bibunit}


\begin{thebibliography}{10}
\renewcommand*{\sfdefault}{PTSansNarrow-TLF}

\bibitem{anstine_generative_2023}
\href{https://doi.org/10.1021/jacs.2c13467}{D.~M. Anstine and O.~Isayev}.
\newblock \href{https://doi.org/10.1021/jacs.2c13467}{Generative {Models} as an {Emerging} {Paradigm} in the {Chemical} {Sciences}}.
\newblock \href{https://doi.org/10.1021/jacs.2c13467}{{\em Journal of the American Chemical Society}}, \href{https://doi.org/10.1021/jacs.2c13467}{145(16):8736--8750}, \href{https://doi.org/10.1021/jacs.2c13467}{2023}. \href{https://doi.org/10.1021/jacs.2c13467}
{doi: \textsf{%
10\hspace{.1pt}\discretionary{.}{%
}{.}\hspace{.4pt}1021\discretionary{/}{%
}{/}jacs\hspace{.1pt}\discretionary{.}{%
}{.}\hspace{.4pt}2c13467}}


\bibitem{bienfait_jsme_2013}
\href{https://doi.org/10.1186/1758-2946-5-24}{B.~Bienfait and P.~Ertl}.
\newblock \href{https://doi.org/10.1186/1758-2946-5-24}{{JSME}: a free molecule editor in {JavaScript}}.
\newblock \href{https://doi.org/10.1186/1758-2946-5-24}{{\em Journal of Cheminformatics}}, \href{https://doi.org/10.1186/1758-2946-5-24}{5(1):24}, \href{https://doi.org/10.1186/1758-2946-5-24}{2013}. \href{https://doi.org/10.1186/1758-2946-5-24}
{doi: \textsf{%
10\hspace{.1pt}\discretionary{.}{%
}{.}\hspace{.4pt}1186\discretionary{/}{%
}{/}1758\discretionary{%
}{-}{-}2946\discretionary{%
}{-}{-}5\discretionary{%
}{-}{-}24}}


\bibitem{brown_guacamol_2019}
\href{https://doi.org/10.1021/acs.jcim.8b00839}{N.~Brown, M.~Fiscato, M.~H. Segler, and A.~C. Vaucher}.
\newblock \href{https://doi.org/10.1021/acs.jcim.8b00839}{{GuacaMol}: {Benchmarking} {Models} for de {Novo} {Molecular} {Design}}.
\newblock \href{https://doi.org/10.1021/acs.jcim.8b00839}{{\em Journal of Chemical Information and Modeling}}, \href{https://doi.org/10.1021/acs.jcim.8b00839}{59(3):1096--1108}, \href{https://doi.org/10.1021/acs.jcim.8b00839}{2019}. \href{https://doi.org/10.1021/acs.jcim.8b00839}
{doi: \textsf{%
10\hspace{.1pt}\discretionary{.}{%
}{.}\hspace{.4pt}1021\discretionary{/}{%
}{/}acs\hspace{.1pt}\discretionary{.}{%
}{.}\hspace{.4pt}jcim\hspace{.1pt}\discretionary{.}{%
}{.}\hspace{.4pt}8b00839}}


\bibitem{brunn_pfas_2023}
\href{https://doi.org/10.1186/s12302-023-00721-8}{H.~Brunn, G.~Arnold, W.~Körner, G.~Rippen, K.~G. Steinhäuser, and I.~Valentin}.
\newblock \href{https://doi.org/10.1186/s12302-023-00721-8}{{PFAS}: forever chemicals—persistent, bioaccumulative and mobile. {Reviewing} the status and the need for their phase out and remediation of contaminated sites}.
\newblock \href{https://doi.org/10.1186/s12302-023-00721-8}{{\em Environmental Sciences Europe}}, \href{https://doi.org/10.1186/s12302-023-00721-8}{35(1):20}, \href{https://doi.org/10.1186/s12302-023-00721-8}{2023}. \href{https://doi.org/10.1186/s12302-023-00721-8}
{doi: \textsf{%
10\hspace{.1pt}\discretionary{.}{%
}{.}\hspace{.4pt}1186\discretionary{/}{%
}{/}s12302\discretionary{%
}{-}{-}023\discretionary{%
}{-}{-}00721\discretionary{%
}{-}{-}8}}


\bibitem{choung_extracting_2023}
\href{https://doi.org/10.1038/s41467-023-42242-1}{O.-H. Choung, R.~Vianello, M.~Segler, N.~Stiefl, and J.~Jiménez-Luna}.
\newblock \href{https://doi.org/10.1038/s41467-023-42242-1}{Extracting medicinal chemistry intuition via preference machine learning}.
\newblock \href{https://doi.org/10.1038/s41467-023-42242-1}{{\em Nature Communications}}, \href{https://doi.org/10.1038/s41467-023-42242-1}{14(1):6651}, \href{https://doi.org/10.1038/s41467-023-42242-1}{2023}. \href{https://doi.org/10.1038/s41467-023-42242-1}
{doi: \textsf{%
10\hspace{.1pt}\discretionary{.}{%
}{.}\hspace{.4pt}1038\discretionary{/}{%
}{/}s41467\discretionary{%
}{-}{-}023\discretionary{%
}{-}{-}42242\discretionary{%
}{-}{-}1}}


\bibitem{cihan_sorkun_chemplot_2022}
\href{https://doi.org/10.1002/cmtd.202200005}{M.~Cihan~Sorkun, D.~Mullaj, J.~M. V.~A. Koelman, and S.~Er}.
\newblock \href{https://doi.org/10.1002/cmtd.202200005}{{ChemPlot}, a {Python} {Library} for {Chemical} {Space} {Visualization}}.
\newblock \href{https://doi.org/10.1002/cmtd.202200005}{{\em Chemistry–Methods}}, \href{https://doi.org/10.1002/cmtd.202200005}{2(7):e202200005}, \href{https://doi.org/10.1002/cmtd.202200005}{2022}. \href{https://doi.org/10.1002/cmtd.202200005}
{doi: \textsf{%
10\hspace{.1pt}\discretionary{.}{%
}{.}\hspace{.4pt}1002\discretionary{/}{%
}{/}cmtd\hspace{.1pt}\discretionary{.}{%
}{.}\hspace{.4pt}202200005}}


\bibitem{fenner_need_2021}
\href{https://doi.org/10.1021/acs.est.1c04903}{K.~Fenner and M.~Scheringer}.
\newblock \href{https://doi.org/10.1021/acs.est.1c04903}{The {Need} for {Chemical} {Simplification} {As} a {Logical} {Consequence} of {Ever}-{Increasing} {Chemical} {Pollution}}.
\newblock \href{https://doi.org/10.1021/acs.est.1c04903}{{\em Environmental Science \& Technology}}, \href{https://doi.org/10.1021/acs.est.1c04903}{55(21):14470--14472}, \href{https://doi.org/10.1021/acs.est.1c04903}{2021}. \href{https://doi.org/10.1021/acs.est.1c04903}
{doi: \textsf{%
10\hspace{.1pt}\discretionary{.}{%
}{.}\hspace{.4pt}1021\discretionary{/}{%
}{/}acs\hspace{.1pt}\discretionary{.}{%
}{.}\hspace{.4pt}est\hspace{.1pt}\discretionary{.}{%
}{.}\hspace{.4pt}1c04903}}


\bibitem{fujii_chemtsv3_2025}
\href{https://doi.org/10.26434/chemrxiv-2025-kdvrt}{S.~Fujii, Y.~Murakami, T.~Yoshizawa, S.~Ishida, N.~Cho, M.~Ohta, T.~Honma, K.~Yoshizoe, M.~Sumita, K.~Tsuda, and K.~Terayama}.
\newblock \href{https://doi.org/10.26434/chemrxiv-2025-kdvrt}{{ChemTSv3}: {Generalizing} {Molecular} {Design} via {Flexible} {Search} {Space} {Control}}.
\newblock \href{https://doi.org/10.26434/chemrxiv-2025-kdvrt}{2025}. \href{https://doi.org/10.26434/chemrxiv-2025-kdvrt}
{doi: \textsf{%
10\hspace{.1pt}\discretionary{.}{%
}{.}\hspace{.4pt}26434\discretionary{/}{%
}{/}chemrxiv\discretionary{%
}{-}{-}2025\discretionary{%
}{-}{-}kdvrt}}


\bibitem{goldszal_discovery_2025}
\href{https://doi.org/10.48550/arXiv.2509.19588}{A.~Goldszal, D.~Calanzone, V.~Taboga, and P.-L. Bacon}.
\newblock \href{https://doi.org/10.48550/arXiv.2509.19588}{Discovery of {Sustainable} {Refrigerants} through {Physics}-{Informed} {RL} {Fine}-{Tuning} of {Sequence} {Models}}, \href{https://doi.org/10.48550/arXiv.2509.19588}{2025}. \href{https://doi.org/10.48550/arXiv.2509.19588}
{doi: \textsf{%
10\hspace{.1pt}\discretionary{.}{%
}{.}\hspace{.4pt}48550\discretionary{/}{%
}{/}arXiv\hspace{.1pt}\discretionary{.}{%
}{.}\hspace{.4pt}2509\hspace{.1pt}\discretionary{.}{%
}{.}\hspace{.4pt}19588}}


\bibitem{gratzl_lineup_2013}
\href{https://doi.org/10.1109/TVCG.2013.173}{S.~Gratzl, A.~Lex, N.~Gehlenborg, H.~Pfister, and M.~Streit}.
\newblock \href{https://doi.org/10.1109/TVCG.2013.173}{{LineUp}: {Visual} {Analysis} of {Multi}-{Attribute} {Rankings}}.
\newblock \href{https://doi.org/10.1109/TVCG.2013.173}{{\em {IEEE} Transactions on Visualization and Computer Graphics}}, \href{https://doi.org/10.1109/TVCG.2013.173}{19(12):2277--2286}, \href{https://doi.org/10.1109/TVCG.2013.173}{2013}. \href{https://doi.org/10.1109/TVCG.2013.173}
{doi: \textsf{%
10\hspace{.1pt}\discretionary{.}{%
}{.}\hspace{.4pt}1109\discretionary{/}{%
}{/}TVCG\hspace{.1pt}\discretionary{.}{%
}{.}\hspace{.4pt}2013\hspace{.1pt}\discretionary{.}{%
}{.}\hspace{.4pt}173}}


\bibitem{he_collaborative_2025}
\href{https://doi.org/10.1287/isre.2024.1154}{J.~He, C.~Hua, Y.~Wang, and Z.~Zheng}.
\newblock \href{https://doi.org/10.1287/isre.2024.1154}{Collaborative {Intelligence} in {Sequential} {Experiments}: {A} {Human}-in-the-{Loop} {Framework} for {Drug} {Discovery}}.
\newblock \href{https://doi.org/10.1287/isre.2024.1154}{{\em Information Systems Research}}, \href{https://doi.org/10.1287/isre.2024.1154}{2025}. \href{https://doi.org/10.1287/isre.2024.1154}
{doi: \textsf{%
10\hspace{.1pt}\discretionary{.}{%
}{.}\hspace{.4pt}1287\discretionary{/}{%
}{/}isre\hspace{.1pt}\discretionary{.}{%
}{.}\hspace{.4pt}2024\hspace{.1pt}\discretionary{.}{%
}{.}\hspace{.4pt}1154}}


\bibitem{humer_cheminformatics_2022}
\href{https://doi.org/10.1186/s13321-022-00600-z}{C.~Humer, H.~Heberle, F.~Montanari, T.~Wolf, F.~Huber, R.~Henderson, J.~Heinrich, and M.~Streit}.
\newblock \href{https://doi.org/10.1186/s13321-022-00600-z}{{ChemInformatics} {Model} {Explorer} ({CIME}): exploratory analysis of chemical model explanations}.
\newblock \href{https://doi.org/10.1186/s13321-022-00600-z}{{\em Journal of Cheminformatics}}, \href{https://doi.org/10.1186/s13321-022-00600-z}{14(1):21}, \href{https://doi.org/10.1186/s13321-022-00600-z}{2022}. \href{https://doi.org/10.1186/s13321-022-00600-z}
{doi: \textsf{%
10\hspace{.1pt}\discretionary{.}{%
}{.}\hspace{.4pt}1186\discretionary{/}{%
}{/}s13321\discretionary{%
}{-}{-}022\discretionary{%
}{-}{-}00600\discretionary{%
}{-}{-}z}}


\bibitem{humer_cime4r_2024}
\href{https://doi.org/10.1186/s13321-024-00840-1}{C.~Humer, R.~Nicholls, H.~Heberle, M.~Heckmann, M.~Pühringer, T.~Wolf, M.~Lübbesmeyer, J.~Heinrich, J.~Hillenbrand, G.~Volpin, and M.~Streit}.
\newblock \href{https://doi.org/10.1186/s13321-024-00840-1}{{CIME4R}: {Exploring} iterative, {AI}-guided chemical reaction optimization campaigns in their parameter space}.
\newblock \href{https://doi.org/10.1186/s13321-024-00840-1}{{\em Journal of Cheminformatics}}, \href{https://doi.org/10.1186/s13321-024-00840-1}{16(1):51}, \href{https://doi.org/10.1186/s13321-024-00840-1}{2024}. \href{https://doi.org/10.1186/s13321-024-00840-1}
{doi: \textsf{%
10\hspace{.1pt}\discretionary{.}{%
}{.}\hspace{.4pt}1186\discretionary{/}{%
}{/}s13321\discretionary{%
}{-}{-}024\discretionary{%
}{-}{-}00840\discretionary{%
}{-}{-}1}}


\bibitem{jain_gflownets_2023}
\href{https://doi.org/10.1039/D3DD00002H}{M.~Jain, T.~Deleu, J.~Hartford, C.-H. Liu, A.~Hernandez-Garcia, and Y.~Bengio}.
\newblock \href{https://doi.org/10.1039/D3DD00002H}{{GFlowNets} for {AI}-driven scientific discovery}.
\newblock \href{https://doi.org/10.1039/D3DD00002H}{{\em Digital Discovery}}, \href{https://doi.org/10.1039/D3DD00002H}{2(3):557--577}, \href{https://doi.org/10.1039/D3DD00002H}{2023}. \href{https://doi.org/10.1039/D3DD00002H}
{doi: \textsf{%
10\hspace{.1pt}\discretionary{.}{%
}{.}\hspace{.4pt}1039\discretionary{/}{%
}{/}D3DD00002H}}


\bibitem{jayasekara_assessing_2021}
\href{https://doi.org/10.1016/j.yrtph.2021.105006}{P.~S. Jayasekara, S.~K. Skanchy, M.~T. Kim, G.~Kumaran, B.~E. Mugabe, L.~E. Woodard, J.~Yang, A.~J. Zych, and N.~L. Kruhlak}.
\newblock \href{https://doi.org/10.1016/j.yrtph.2021.105006}{Assessing the impact of expert knowledge on {ICH} {M7} ({Q}){SAR} predictions. {Is} expert review still needed?}
\newblock \href{https://doi.org/10.1016/j.yrtph.2021.105006}{{\em Regulatory Toxicology and Pharmacology}}, \href{https://doi.org/10.1016/j.yrtph.2021.105006}{125:105006}, \href{https://doi.org/10.1016/j.yrtph.2021.105006}{Oct. 2021}. \href{https://doi.org/10.1016/j.yrtph.2021.105006}
{doi: \textsf{%
10\hspace{.1pt}\discretionary{.}{%
}{.}\hspace{.4pt}1016\discretionary{/}{%
}{/}j\hspace{.1pt}\discretionary{.}{%
}{.}\hspace{.4pt}yrtph\hspace{.1pt}\discretionary{.}{%
}{.}\hspace{.4pt}2021\hspace{.1pt}\discretionary{.}{%
}{.}\hspace{.4pt}105006}}


\bibitem{jensen_graph-based_2019}
\href{https://doi.org/10.1039/C8SC05372C}{J.~H. Jensen}.
\newblock \href{https://doi.org/10.1039/C8SC05372C}{A graph-based genetic algorithm and generative model/{Monte} {Carlo} tree search for the exploration of chemical space}.
\newblock \href{https://doi.org/10.1039/C8SC05372C}{{\em Chemical Science}}, \href{https://doi.org/10.1039/C8SC05372C}{10(12):3567--3572}, \href{https://doi.org/10.1039/C8SC05372C}{2019}. \href{https://doi.org/10.1039/C8SC05372C}
{doi: \textsf{%
10\hspace{.1pt}\discretionary{.}{%
}{.}\hspace{.4pt}1039\discretionary{/}{%
}{/}C8SC05372C}}


\bibitem{kale_chemograph_2023}
\href{https://doi.org/10.1111/cgf.14807}{B.~Kale, A.~Clyde, M.~Sun, A.~Ramanathan, R.~Stevens, and M.~E. Papka}.
\newblock \href{https://doi.org/10.1111/cgf.14807}{{ChemoGraph}: {Interactive} {Visual} {Exploration} of the {Chemical} {Space}}.
\newblock \href{https://doi.org/10.1111/cgf.14807}{{\em Computer Graphics Forum}}, \href{https://doi.org/10.1111/cgf.14807}{42(3):13--24}, \href{https://doi.org/10.1111/cgf.14807}{2023}. \href{https://doi.org/10.1111/cgf.14807}
{doi: \textsf{%
10\hspace{.1pt}\discretionary{.}{%
}{.}\hspace{.4pt}1111\discretionary{/}{%
}{/}cgf\hspace{.1pt}\discretionary{.}{%
}{.}\hspace{.4pt}14807}}


\bibitem{menke_metis_2024}
\href{https://doi.org/10.1186/s13321-024-00892-3}{J.~Menke, Y.~Nahal, E.~J. Bjerrum, M.~Kabeshov, S.~Kaski, and O.~Engkvist}.
\newblock \href{https://doi.org/10.1186/s13321-024-00892-3}{Metis: a python-based user interface to collect expert feedback for generative chemistry models}.
\newblock \href{https://doi.org/10.1186/s13321-024-00892-3}{{\em Journal of Cheminformatics}}, \href{https://doi.org/10.1186/s13321-024-00892-3}{16(1):100}, \href{https://doi.org/10.1186/s13321-024-00892-3}{2024}. \href{https://doi.org/10.1186/s13321-024-00892-3}
{doi: \textsf{%
10\hspace{.1pt}\discretionary{.}{%
}{.}\hspace{.4pt}1186\discretionary{/}{%
}{/}s13321\discretionary{%
}{-}{-}024\discretionary{%
}{-}{-}00892\discretionary{%
}{-}{-}3}}


\bibitem{mitchell_introduction_1996}
M.~Mitchell.
\newblock {\em An introduction to genetic algorithms}.
\newblock Complex adaptive systems. MIT Press, 1996.

\bibitem{mosqueira-rey_human---loop_2023}
\href{https://doi.org/10.1007/s10462-022-10246-w}{E.~Mosqueira-Rey, E.~Hernández-Pereira, D.~Alonso-Ríos, J.~Bobes-Bascarán, and {\'A}.~Fernández-Leal}.
\newblock \href{https://doi.org/10.1007/s10462-022-10246-w}{Human-in-the-loop machine learning: a state of the art}.
\newblock \href{https://doi.org/10.1007/s10462-022-10246-w}{{\em Artificial Intelligence Review}}, \href{https://doi.org/10.1007/s10462-022-10246-w}{56(4):3005--3054}, \href{https://doi.org/10.1007/s10462-022-10246-w}{2023}. \href{https://doi.org/10.1007/s10462-022-10246-w}
{doi: \textsf{%
10\hspace{.1pt}\discretionary{.}{%
}{.}\hspace{.4pt}1007\discretionary{/}{%
}{/}s10462\discretionary{%
}{-}{-}022\discretionary{%
}{-}{-}10246\discretionary{%
}{-}{-}w}}


\bibitem{nahal_human---loop_2024}
\href{https://doi.org/10.1186/s13321-024-00924-y}{Y.~Nahal, J.~Menke, J.~Martinelli, M.~Heinonen, M.~Kabeshov, J.~P. Janet, E.~Nittinger, O.~Engkvist, and S.~Kaski}.
\newblock \href{https://doi.org/10.1186/s13321-024-00924-y}{Human-in-the-loop active learning for goal-oriented molecule generation}.
\newblock \href{https://doi.org/10.1186/s13321-024-00924-y}{{\em Journal of Cheminformatics}}, \href{https://doi.org/10.1186/s13321-024-00924-y}{16(1):138}, \href{https://doi.org/10.1186/s13321-024-00924-y}{2024}. \href{https://doi.org/10.1186/s13321-024-00924-y}
{doi: \textsf{%
10\hspace{.1pt}\discretionary{.}{%
}{.}\hspace{.4pt}1186\discretionary{/}{%
}{/}s13321\discretionary{%
}{-}{-}024\discretionary{%
}{-}{-}00924\discretionary{%
}{-}{-}y}}


\bibitem{posthumus_external_2005}
\href{https://doi.org/10.1080/10629360412331319899}{R.~Posthumus, T.~Traas, W.~Peijnenburg, and E.~Hulzebos}.
\newblock \href{https://doi.org/10.1080/10629360412331319899}{External validation of {EPIWIN} biodegradation models}.
\newblock \href{https://doi.org/10.1080/10629360412331319899}{{\em SAR and QSAR in Environmental Research}}, \href{https://doi.org/10.1080/10629360412331319899}{16(1-2):135--148}, \href{https://doi.org/10.1080/10629360412331319899}{Feb. 2005}. \href{https://doi.org/10.1080/10629360412331319899}
{doi: \textsf{%
10\hspace{.1pt}\discretionary{.}{%
}{.}\hspace{.4pt}1080\discretionary{/}{%
}{/}10629360412331319899}}


\bibitem{rodriguez-perez_interpretation_2020}
\href{https://doi.org/10.1021/acs.jmedchem.9b01101}{R.~Rodríguez-Pérez and J.~Bajorath}.
\newblock \href{https://doi.org/10.1021/acs.jmedchem.9b01101}{Interpretation of {Compound} {Activity} {Predictions} from {Complex} {Machine} {Learning} {Models} {Using} {Local} {Approximations} and {Shapley} {Values}}.
\newblock \href{https://doi.org/10.1021/acs.jmedchem.9b01101}{{\em Journal of Medicinal Chemistry}}, \href{https://doi.org/10.1021/acs.jmedchem.9b01101}{63(16):8761--8777}, \href{https://doi.org/10.1021/acs.jmedchem.9b01101}{2020}. \href{https://doi.org/10.1021/acs.jmedchem.9b01101}
{doi: \textsf{%
10\hspace{.1pt}\discretionary{.}{%
}{.}\hspace{.4pt}1021\discretionary{/}{%
}{/}acs\hspace{.1pt}\discretionary{.}{%
}{.}\hspace{.4pt}jmedchem\hspace{.1pt}\discretionary{.}{%
}{.}\hspace{.4pt}9b01101}}


\bibitem{rogers_extended-connectivity_2010}
\href{https://doi.org/10.1021/ci100050t}{D.~Rogers and M.~Hahn}.
\newblock \href{https://doi.org/10.1021/ci100050t}{Extended-{Connectivity} {Fingerprints}}.
\newblock \href{https://doi.org/10.1021/ci100050t}{{\em Journal of Chemical Information and Modeling}}, \href{https://doi.org/10.1021/ci100050t}{50(5):742--754}, \href{https://doi.org/10.1021/ci100050t}{2010}. \href{https://doi.org/10.1021/ci100050t}
{doi: \textsf{%
10\hspace{.1pt}\discretionary{.}{%
}{.}\hspace{.4pt}1021\discretionary{/}{%
}{/}ci100050t}}


\bibitem{rubin_bisphenol_2011}
\href{https://doi.org/10.1016/j.jsbmb.2011.05.002}{B.~S. Rubin}.
\newblock \href{https://doi.org/10.1016/j.jsbmb.2011.05.002}{Bisphenol {A}: {An} endocrine disruptor with widespread exposure and multiple effects}.
\newblock \href{https://doi.org/10.1016/j.jsbmb.2011.05.002}{{\em The Journal of Steroid Biochemistry and Molecular Biology}}, \href{https://doi.org/10.1016/j.jsbmb.2011.05.002}{127(1-2):27--34}, \href{https://doi.org/10.1016/j.jsbmb.2011.05.002}{Oct. 2011}. \href{https://doi.org/10.1016/j.jsbmb.2011.05.002}
{doi: \textsf{%
10\hspace{.1pt}\discretionary{.}{%
}{.}\hspace{.4pt}1016\discretionary{/}{%
}{/}j\hspace{.1pt}\discretionary{.}{%
}{.}\hspace{.4pt}jsbmb\hspace{.1pt}\discretionary{.}{%
}{.}\hspace{.4pt}2011\hspace{.1pt}\discretionary{.}{%
}{.}\hspace{.4pt}05\hspace{.1pt}\discretionary{.}{%
}{.}\hspace{.4pt}002}}


\bibitem{seller-brison_hazard_2026}
\href{https://doi.org/10.1021/acs.estlett.5c01217}{C.~Seller-Brison, F.~Weissbach, K.~Jorner, M.~Scheringer, and K.~Fenner}.
\newblock \href{https://doi.org/10.1021/acs.estlett.5c01217}{Hazard {Assessment} of {Antioxidants} as {Contaminants} of {Concern}}.
\newblock \href{https://doi.org/10.1021/acs.estlett.5c01217}{{\em Environmental Science \& Technology Letters}}, \href{https://doi.org/10.1021/acs.estlett.5c01217}{2026}. \href{https://doi.org/10.1021/acs.estlett.5c01217}
{doi: \textsf{%
10\hspace{.1pt}\discretionary{.}{%
}{.}\hspace{.4pt}1021\discretionary{/}{%
}{/}acs\hspace{.1pt}\discretionary{.}{%
}{.}\hspace{.4pt}estlett\hspace{.1pt}\discretionary{.}{%
}{.}\hspace{.4pt}5c01217}}


\bibitem{sundin_human---loop_2022}
\href{https://doi.org/10.1186/s13321-022-00667-8}{I.~Sundin, A.~Voronov, H.~Xiao, K.~Papadopoulos, E.~J. Bjerrum, M.~Heinonen, A.~Patronov, S.~Kaski, and O.~Engkvist}.
\newblock \href{https://doi.org/10.1186/s13321-022-00667-8}{Human-in-the-loop assisted de novo molecular design}.
\newblock \href{https://doi.org/10.1186/s13321-022-00667-8}{{\em Journal of Cheminformatics}}, \href{https://doi.org/10.1186/s13321-022-00667-8}{14(1):86}, \href{https://doi.org/10.1186/s13321-022-00667-8}{2022}. \href{https://doi.org/10.1186/s13321-022-00667-8}
{doi: \textsf{%
10\hspace{.1pt}\discretionary{.}{%
}{.}\hspace{.4pt}1186\discretionary{/}{%
}{/}s13321\discretionary{%
}{-}{-}022\discretionary{%
}{-}{-}00667\discretionary{%
}{-}{-}8}}


\bibitem{tripp_genetic_2023}
\href{https://doi.org/10.48550/ARXIV.2310.09267}{A.~Tripp and J.~M. Hernández-Lobato}.
\newblock \href{https://doi.org/10.48550/ARXIV.2310.09267}{Genetic algorithms are strong baselines for molecule generation}, \href{https://doi.org/10.48550/ARXIV.2310.09267}{2023}. \href{https://doi.org/10.48550/ARXIV.2310.09267}
{doi: \textsf{%
10\hspace{.1pt}\discretionary{.}{%
}{.}\hspace{.4pt}48550\discretionary{/}{%
}{/}ARXIV\hspace{.1pt}\discretionary{.}{%
}{.}\hspace{.4pt}2310\hspace{.1pt}\discretionary{.}{%
}{.}\hspace{.4pt}09267}}


\bibitem{vieira_imesc_2025}
\href{https://doi.org/10.3389/fenvs.2025.1533292}{D.~C. Vieira, F.~S. Paula, L.~E. Yaginuma, and G.~Fonseca}.
\newblock \href{https://doi.org/10.3389/fenvs.2025.1533292}{{iMESc} – an interactive machine learning app for environmental sciences}.
\newblock \href{https://doi.org/10.3389/fenvs.2025.1533292}{{\em Frontiers in Environmental Science}}, \href{https://doi.org/10.3389/fenvs.2025.1533292}{13}, \href{https://doi.org/10.3389/fenvs.2025.1533292}{2025}. \href{https://doi.org/10.3389/fenvs.2025.1533292}
{doi: \textsf{%
10\hspace{.1pt}\discretionary{.}{%
}{.}\hspace{.4pt}3389\discretionary{/}{%
}{/}fenvs\hspace{.1pt}\discretionary{.}{%
}{.}\hspace{.4pt}2025\hspace{.1pt}\discretionary{.}{%
}{.}\hspace{.4pt}1533292}}


\bibitem{wang_paretogen_2026}
\href{https://doi.org/10.1021/acs.est.6c00350}{H.~Wang, W.~Liu, J.~Chen, S.~Ji, and G.~Bi}.
\newblock \href{https://doi.org/10.1021/acs.est.6c00350}{{ParetoGen}: {Generative} {Machine} {Learning} {Models} {To} {Push} the {Pareto} {Optimal} {Frontier} of {Functionality}-{Hazard} {Trade}-offs in {Per}- and {Polyfluoroalkyl} {Substances} {Green} {Alternative} {Designs}}.
\newblock \href{https://doi.org/10.1021/acs.est.6c00350}{{\em Environmental Science \& Technology}}, \href{https://doi.org/10.1021/acs.est.6c00350}{2026}. \href{https://doi.org/10.1021/acs.est.6c00350}
{doi: \textsf{%
10\hspace{.1pt}\discretionary{.}{%
}{.}\hspace{.4pt}1021\discretionary{/}{%
}{/}acs\hspace{.1pt}\discretionary{.}{%
}{.}\hspace{.4pt}est\hspace{.1pt}\discretionary{.}{%
}{.}\hspace{.4pt}6c00350}}


\bibitem{wang_efficient_2025}
\href{https://openreview.net/forum?id=awWiNvQwf3}{H.~Wang, M.~Skreta, C.~T. Ser, W.~Gao, L.~Kong, F.~Strieth-Kalthoff, C.~Duan, Y.~Zhuang, Y.~Yu, Y.~Zhu, Y.~Du, A.~Aspuru-Guzik, K.~Neklyudov, and C.~Zhang}.
\newblock \href{https://openreview.net/forum?id=awWiNvQwf3}{Efficient {Evolutionary} {Search} {Over} {Chemical} {Space} with {Large} {Language} {Models}}.
\newblock \href{https://openreview.net/forum?id=awWiNvQwf3}{In {\em International Conference on Learning Representations}}, \href{https://openreview.net/forum?id=awWiNvQwf3}{2025}.

\bibitem{wang_addressing_2025}
\href{https://doi.org/10.1021/acs.est.5c00510}{Y.~Wang, J.~Dong, Y.~Zhou, Y.~Cheng, X.~Zhao, W.~J. G.~M. Peijnenburg, M.~G. Vijver, K.~M.~Y. Leung, W.~Fan, and F.~Wu}.
\newblock \href{https://doi.org/10.1021/acs.est.5c00510}{Addressing the {Data} {Scarcity} {Problem} in {Ecotoxicology} via {Small} {Data} {Machine} {Learning} {Methods}}.
\newblock \href{https://doi.org/10.1021/acs.est.5c00510}{{\em Environmental Science \& Technology}}, \href{https://doi.org/10.1021/acs.est.5c00510}{59(12):5867--5871}, \href{https://doi.org/10.1021/acs.est.5c00510}{2025}. \href{https://doi.org/10.1021/acs.est.5c00510}
{doi: \textsf{%
10\hspace{.1pt}\discretionary{.}{%
}{.}\hspace{.4pt}1021\discretionary{/}{%
}{/}acs\hspace{.1pt}\discretionary{.}{%
}{.}\hspace{.4pt}est\hspace{.1pt}\discretionary{.}{%
}{.}\hspace{.4pt}5c00510}}


\bibitem{yang_molecular_2024}
\href{https://doi.org/10.1016/j.scitotenv.2024.176095}{Y.~Yang, Z.~Yang, X.~Pang, H.~Cao, Y.~Sun, L.~Wang, Z.~Zhou, P.~Wang, Y.~Liang, and Y.~Wang}.
\newblock \href{https://doi.org/10.1016/j.scitotenv.2024.176095}{Molecular designing of potential environmentally friendly {PFAS} based on deep learning and generative models}.
\newblock \href{https://doi.org/10.1016/j.scitotenv.2024.176095}{{\em Science of The Total Environment}}, \href{https://doi.org/10.1016/j.scitotenv.2024.176095}{953:176095}, \href{https://doi.org/10.1016/j.scitotenv.2024.176095}{2024}. \href{https://doi.org/10.1016/j.scitotenv.2024.176095}
{doi: \textsf{%
10\hspace{.1pt}\discretionary{.}{%
}{.}\hspace{.4pt}1016\discretionary{/}{%
}{/}j\hspace{.1pt}\discretionary{.}{%
}{.}\hspace{.4pt}scitotenv\hspace{.1pt}\discretionary{.}{%
}{.}\hspace{.4pt}2024\hspace{.1pt}\discretionary{.}{%
}{.}\hspace{.4pt}176095}}


\end{thebibliography}


\begin{thebibliography}{1}
\renewcommand*{\sfdefault}{PTSansNarrow-TLF}

\bibitem{RDKitOpensourceCheminformatics}
\href{https://www.rdkit.org/}{{RDKit}: {Open}-source cheminformatics}.

\bibitem{bemis_1996_PropertiesKnownDrugs}
\href{https://doi.org/10.1021/jm9602928}{G.~W. Bemis and M.~A. Murcko}.
\newblock \href{https://doi.org/10.1021/jm9602928}{The {Properties} of {Known} {Drugs}. 1. {Molecular} {Frameworks}}.
\newblock \href{https://doi.org/10.1021/jm9602928}{{\em Journal of Medicinal Chemistry}}, \href{https://doi.org/10.1021/jm9602928}{39(15):2887--2893}, \href{https://doi.org/10.1021/jm9602928}{Jan. 1996}. \href{https://doi.org/10.1021/jm9602928}
{doi: \textsf{%
10\hspace{.1pt}\discretionary{.}{%
}{.}\hspace{.4pt}1021\discretionary{/}{%
}{/}jm9602928}}


\bibitem{ertl_2009_EstimationSyntheticAccessibility}
\href{https://doi.org/10.1186/1758-2946-1-8}{P.~Ertl and A.~Schuffenhauer}.
\newblock \href{https://doi.org/10.1186/1758-2946-1-8}{Estimation of synthetic accessibility score of drug-like molecules based on molecular complexity and fragment contributions}.
\newblock \href{https://doi.org/10.1186/1758-2946-1-8}{{\em Journal of Cheminformatics}}, \href{https://doi.org/10.1186/1758-2946-1-8}{1(1):1--11}, \href{https://doi.org/10.1186/1758-2946-1-8}{Dec. 2009}.
\newblock \href{https://doi.org/10.1186/1758-2946-1-8}{Number: 1}. \href{https://doi.org/10.1186/1758-2946-1-8}
{doi: \textsf{%
10\hspace{.1pt}\discretionary{.}{%
}{.}\hspace{.4pt}1186\discretionary{/}{%
}{/}1758\discretionary{%
}{-}{-}2946\discretionary{%
}{-}{-}1\discretionary{%
}{-}{-}8}}


\bibitem{genheden_aizynthfinder_2020}
\href{https://doi.org/10.1186/s13321-020-00472-1}{S.~Genheden, A.~Thakkar, V.~Chadimová, J.-L. Reymond, O.~Engkvist, and E.~Bjerrum}.
\newblock \href{https://doi.org/10.1186/s13321-020-00472-1}{{AiZynthFinder}: a fast, robust and flexible open-source software for retrosynthetic planning}.
\newblock \href{https://doi.org/10.1186/s13321-020-00472-1}{{\em Journal of Cheminformatics}}, \href{https://doi.org/10.1186/s13321-020-00472-1}{12(1):70}, \href{https://doi.org/10.1186/s13321-020-00472-1}{Dec. 2020}. \href{https://doi.org/10.1186/s13321-020-00472-1}
{doi: \textsf{%
10\hspace{.1pt}\discretionary{.}{%
}{.}\hspace{.4pt}1186\discretionary{/}{%
}{/}s13321\discretionary{%
}{-}{-}020\discretionary{%
}{-}{-}00472\discretionary{%
}{-}{-}1}}


\bibitem{jensen_graph-based_2019}
\href{https://doi.org/10.1039/C8SC05372C}{J.~H. Jensen}.
\newblock \href{https://doi.org/10.1039/C8SC05372C}{A graph-based genetic algorithm and generative model/{Monte} {Carlo} tree search for the exploration of chemical space}.
\newblock \href{https://doi.org/10.1039/C8SC05372C}{{\em Chemical Science}}, \href{https://doi.org/10.1039/C8SC05372C}{10(12):3567--3572}, \href{https://doi.org/10.1039/C8SC05372C}{2019}. \href{https://doi.org/10.1039/C8SC05372C}
{doi: \textsf{%
10\hspace{.1pt}\discretionary{.}{%
}{.}\hspace{.4pt}1039\discretionary{/}{%
}{/}C8SC05372C}}


\bibitem{mirza_2025_FrameworkEvaluatingChemical}
\href{https://doi.org/10.1038/s41557-025-01815-x}{A.~Mirza, N.~Alampara, S.~Kunchapu, M.~Ríos-García, B.~Emoekabu, A.~Krishnan, T.~Gupta, M.~Schilling-Wilhelmi, M.~Okereke, A.~Aneesh, M.~Asgari, J.~Eberhardt, A.~M. Elahi, H.~M. Elbeheiry, M.~V. Gil, C.~Glaubitz, M.~Greiner, C.~T. Holick, T.~Hoffmann, A.~Ibrahim, L.~C. Klepsch, Y.~Köster, F.~A. Kreth, J.~Meyer, S.~Miret, J.~M. Peschel, M.~Ringleb, N.~C. Roesner, J.~Schreiber, U.~S. Schubert, L.~M. Stafast, A.~D.~D. Wonanke, M.~Pieler, P.~Schwaller, and K.~M. Jablonka}.
\newblock \href{https://doi.org/10.1038/s41557-025-01815-x}{A framework for evaluating the chemical knowledge and reasoning abilities of large language models against the expertise of chemists}.
\newblock \href{https://doi.org/10.1038/s41557-025-01815-x}{{\em Nature Chemistry}}, \href{https://doi.org/10.1038/s41557-025-01815-x}{17(7):1027--1034}, \href{https://doi.org/10.1038/s41557-025-01815-x}{July 2025}. \href{https://doi.org/10.1038/s41557-025-01815-x}
{doi: \textsf{%
10\hspace{.1pt}\discretionary{.}{%
}{.}\hspace{.4pt}1038\discretionary{/}{%
}{/}s41557\discretionary{%
}{-}{-}025\discretionary{%
}{-}{-}01815\discretionary{%
}{-}{-}x}}


\bibitem{rogers_extended-connectivity_2010}
\href{https://doi.org/10.1021/ci100050t}{D.~Rogers and M.~Hahn}.
\newblock \href{https://doi.org/10.1021/ci100050t}{Extended-{Connectivity} {Fingerprints}}.
\newblock \href{https://doi.org/10.1021/ci100050t}{{\em Journal of Chemical Information and Modeling}}, \href{https://doi.org/10.1021/ci100050t}{50(5):742--754}, \href{https://doi.org/10.1021/ci100050t}{2010}. \href{https://doi.org/10.1021/ci100050t}
{doi: \textsf{%
10\hspace{.1pt}\discretionary{.}{%
}{.}\hspace{.4pt}1021\discretionary{/}{%
}{/}ci100050t}}


\end{thebibliography}
\end{document}